\documentclass[10pt,preprint,onecolumn]{aastex}
\usepackage{natbib,psfig,amsmath}
\bibpunct{(}{)}{;}{a}{}{,}
\newcommand{\bsi}{\mathbf{s}_i}
\newcommand{\bpi}{\mathbf{p}_i}
\newcommand{\bsj}{\mathbf{s}_j}
\newcommand{\bpj}{\mathbf{p}_j}
\newcommand{\bti}{\mathbf{t}_i}
\newcommand{\bg}{\mathbf{g}}
\newcommand{\bSnet}{\mathbf{S}_{net}}
\newcommand{\bStot}{\mathbf{S}_{tot}}

\newcommand{\bI}{\mathbf{I}}

\newcommand{\bVI}{\mathbf{V}_I}

\newcommand{\fdt}{\frac{\delta t}{\Delta t}}

\shorttitle{Statistical Uncertainties in NIR Array Readouts}
\shortauthors{Vacca \& Cushing}

\begin{document}

\title{Non-Linearity Corrections and Statistical Uncertainties Associated with Near-Infrared Arrays}

\author{William D. Vacca}
\affil{SOFIA-USRA, NASA Ames Research Center, MS 144-2, Moffett Field, CA 94035 and
       Dept. of Astronomy, 601 Campbell Hall, University of California, 
       Berkeley, CA, 94720}
\email{wvacca@mail.arc.nasa.gov}

\and

\author{Michael C. Cushing and John T. Rayner}
\affil{Institute for Astronomy, University of Hawai`i, 2680 Woodlawn Drive, 
       Honolulu, HI 96822}
\email{cushing@ifa.hawaii.edu,rayner@ifa.hawaii.edu}

\begin{abstract}

We  derive general equations for non-linearity corrections and statistical uncertainty (variance) estimates for data acquired with near-infrared detectors employing  correlated double sampling, multiple correlated double sampling (Fowler sampling) and uniformly-spaced continuous readout techniques.  We compare our equation for the variance on each pixel associated with Fowler sampling with  measurements obtained from data taken with the array installed in the near-infrared cross-dispersed spectrograph (SpeX) at the NASA Infrared Telescope Facility and find that it provides an accurate representation of the empirical results. This comparison also reveals that the read noise associated with a single readout of the SpeX array increases with the number of non-destructive reads, $n_r$, as  $n_r^{0.16}$. This implies that the {\em effective} read noise of a stored image decreases as $n_r^{-0.34}$, shallower than the expected rate of $n_r^{-0.5}$. The cause of this read noise behavior is
uncertain, but may be due to heating of the array as a result of the multiple read outs. Such behavior
may be generic to arrays that employ correlated or multiple correlated double sampling readouts.

\end{abstract}

\section{Introduction}

Correlated double sampling (CDS) and multiple correlated double sampling (MCDS, also known as Fowler sampling; Fowler \& Gatley 1990) are common techniques used to read out near-infrared arrays. These techniques are used, for example, for recording data obtained with the arrays in SpeX at the NASA Infrared Telescope Facility \citep{rayner03},  in NIRSPEC at the Keck telescope \citep{mclean98}, in the IRCS at the Subaru Telescope \citep{kobayashi00}, and in ISAAC at the Very Large Telescope \citep{moorwood97}. An alternative readout technique, known as continuous, or ``up-the-ramp" sampling, was used for recording data obtained with the CGS2 spectrograph on the United Kingdom Infrared Telescope \citep{chapman90}. A variant of this technique is the recommended readout scheme for the detectors in the Near-Infrared Camera and Multi-Object Spectrometer (NICMOS) aboard the Hubble Space Telescope \citep{roye03}. The relative advantages and disadvantages of these techniques in various noise-limited operating regimes have been summarized by Garnett \& Forrest (1993).

As part of the effort to develop a set of semi-automated routines to reduce SpeX data \citep[{\em Spextool}; see][]{cushing03}, we have derived equations for the non-linearity corrections and the uncertainties on each pixel value recorded by an array that employs CDS, MCDS, and uniformly-spaced continuous sampling readout techniques. In this paper, we present a simple model 
for the CDS and MCDS readout procedures of a near-infrared array, define the relevant parameters and derive equations for the non-linearity corrections and variances on pixel values associated with these techniques (\S2). The corresponding equations for the uniformly-spaced continuous sampling technique are presented in \S3. Verification of the MCDS equations and their application to the SpeX array
for the determination of the gain and read noise are presented in \S4. A summary of our results is given in \S5.

\section{Correlated and Multiple Correlated Double Sampling}

\subsection{Ideal Detector}
Figure 1a shows a schematic of the signal recorded on an integrating node as a function of time for an ideal detector exposed to a constant source flux and read out using the MCDS technique. The ideal detector is perfectly linear over its entire well-depth, thermally stable, and not subject to any time dependent effects which vary during integrations or readouts (such as bias drift, image persistence that decays over time, or amplifier glow associated with readouts). Excellent descriptions of such systematic effects, which will assumed to be either negligible or correctible (and therefore ignored) in our analysis, are given by Skinner \& Bergeron (1997) and Roye et al.\ (2003) for the NICMOS array.

At time $t=0$, the array is reset to the bias level $\mathbf{b}$, which is assumed to be constant in time during the integrations and readouts.\footnote{Variables written in boldface denote two-dimensional quantities, stored in images, which may have different values at each of the pixels in the image; operations performed on images are carried out on all pixels individually. The individual values at a given pixel $k,l$ are written in italics with subscripts, e.g., $s_{k,l}$.} In near-infrared arrays, the bias is a negative voltage set by the electronics and therefore is ``noise-less''.  Because it is assumed to be simply an additive constant that defines a zero point for count levels at each pixel, it can be removed from all equations without any loss of generality (as long as it does not vary). We assume it takes a time $t_{reset}$ to reset an individual pixel. The reset procedure may be global (all pixels in the array reset at once, as in SpeX) or sequential (array reset occurring pixel-by-pixel, as in NIRSPEC). After reset, the array is immediately read out non-destructively $n_r$ times. The time needed to read an individual pixel is $t_{read}$, and a readout occurs sequentially.
Reading the entire array, therefore, requires a total time $\delta t$, given by the time needed to read out a single pixel times the number of pixels read $n_p$, plus any overhead or waiting time between reads $t_{wait}$,
\begin{equation}
\delta t = n_p t_{read} + t_{wait} ~~~.
\end{equation}
In SpeX, for example, the 1024 $\times$ 1024 array is divided into quadrants, which are read simultaneously and each of which has 8 readout outputs. For this array, $t_{reset} \approx 30 \mu$s, $t_{read} \approx 10 \mu$s, and $n_p = 2^{15}$; therefore $\delta t$ is on the order of 300 ms.

The first set of reads provides an estimate of the mean {\em pedestal} level $\bar{\mathbf{p}}$. After an exposure time $\Delta t$, the array is again read out $n_r$ times, which provides an estimate of the mean {\em signal} level $\bar{\mathbf{s}}$. (In the CDS technique, $n_r = 1$.) Therefore, $\Delta t$ is given by the time difference between the start of the first read of the pedestal $\mathbf{p}_1$ and the start of the first read of the signal $\mathbf{s}_1$. 
In most implementations of these techniques, the values of neither the individual readouts nor the mean pedestal and signal levels are stored; 
instead, usually only the net source counts (in DN) for the $n_r$ non-destructive readouts over an integration time of $\Delta t$, given by,
\begin{eqnarray}
\bSnet & = & \sum_{i=1}^{n_r} \bsi - \sum_{i=1}^{n_r} \bpi \\
             & = & n_r \mathbf{\bar{s}} - n_r \mathbf{\bar{p}} 
\end{eqnarray}
is recorded. The estimated source flux, $\bI$ (which includes contributions from the dark current, background, sky, and object), in units of DN s$^{-1}$ is then given by,
\begin{equation}
\bI = \frac{1}{n_r \Delta t}\bSnet  = \frac{1}{n_r \Delta t}\left (\sum_{i=1}^{n_r} \bsi - \sum_{i=1}^{n_r} \bpi\right )~~.
\end{equation}
Hence, $\bI$ is the slope of the solid diagonal line in Figure 1a.

The goal is to determine the uncertainty, or variance, in the estimated source flux $\bI$ at each pixel.
As can be seen from Equation 4, and will be further demonstrated below (\S2.4), expressions for the individual pedestal and signal readout values are needed for the computation of the variances in $\bI$. From Figure 1a, it can be seen that for a linear detector, after a reset the counts (in DN) recorded at the end of the $i$th read of the pedestal, $\bpi$, and the signal, $\bsi$, can be expressed in terms of the recorded net source counts $\bSnet$ or the estimated source flux $\bI$ as

\begin{eqnarray}
\bpi & = & (i - \mathbf{f})\frac{\bSnet}{n_r}\frac{\delta t}{\Delta t} \\
        & = & (i - \mathbf{f})\bI\delta t ~~.
\end{eqnarray}

\noindent
and
\begin{eqnarray}
\bsi & = & \frac{\bSnet}{n_r} + (i - \mathbf{f})\frac{\bSnet}{n_r}\frac{\delta t}{\Delta t} \\
        & = & \frac{\bSnet}{n_r} + \bpi \\
        & = & \bI\Delta t + \bpi ~~,
\end{eqnarray}

\noindent respectively. Here $\mathbf{f}$ is a fraction, $ \mathbf{f} < 1$, whose value at a given pixel depends on the position of the pixel in the readout sequence. The quantity $(i - f_{k,l})\delta t$ is the time after the reset when the pixel $k,l$ is read out for the $i$th time. If a total number of $n_p$ pixels are read out, $f$ for the $m$th pixel in the readout sequence is given by 
\begin{equation}
f = 1 - m\frac{t_{read}}{\delta t} - \frac{t_{reset}}{\delta t}~~
\end{equation}
for a global reset and 
\begin{equation}
f = 1 - m\frac{t_{read}}{\delta t} - (n_p -  m + 1)\frac{t_{reset}}{\delta t}
\end{equation}
for a sequential reset. 

A common observing technique employed with near-infrared arrays is to immediately repeat an integration $n_c$ times and ``co-add" the results. In this case, the entire array addressing procedure (reset, $n_r$ pedestal reads, $n_r$ signal reads) is repeated $n_c$ times. Usually, the net source counts resulting from the individual integrations are not recorded. Rather, the total source counts given by
\begin{eqnarray}
\bStot = \sum_{j=1}^{n_c} \mathbf{S}_{net,j}  & =  &\sum_{j=1}^{n_c} \sum_{i=1}^{n_r} \mathbf{s}_{i,j} - \sum_{j=1}^{n_c} \sum_{i=1}^{n_r} \mathbf{p}_{i,j} \\
 & = & n_c n_r \mathbf{\bar{s}} - n_c n_r \mathbf{\bar{p}} ~~ .
\end{eqnarray}
\noindent
is stored. The estimated source flux $\bI$ in units of DN s$^{-1}$ is given by,
\begin{equation}
\bI= \frac{1}{n_c n_r \Delta t} \bStot ~~.
\end{equation}

\noindent 

\noindent
Therefore,
\begin{equation}
\bpi = (i - \mathbf{f})\frac{\bStot}{n_cn_r}\frac{\delta t}{\Delta t} ~~,
\end{equation}

\noindent
and
\begin{eqnarray}
\bsi & = & \frac{\bStot}{n_c n_r} + (i - \mathbf{f})\frac{\bStot}{n_c n_r}\frac{\delta t}{\Delta t}  \\
        & = & \frac{\bStot}{n_c n_r} + \bpi ~~.
\end{eqnarray}

\subsection{Non-Linear Detector}

Unfortunately few astronomical detectors are perfectly linear over their entire well depth. Near-infrared arrays are inherently non-linear devices because the detector capacitance 
varies during an integration as photons are detected,  (see e.g., McCaughrean 1988). Figure 1b shows a schematic of the signal recorded on an integrating node as a function of time for a non-linear near-infrared array read out using the MCDS technique. The recorded signal (the solid line) falls below that expected in the ideal case (the diagonal dashed line). In the absence of other systematic and time variable effects (such as image persistence, bias drift, or thermal instability of the array), the ratio of the theoretically-expected  signal to the recorded signal yields the non-linearity curve of the array, which increases steadily from unity as a function of recorded counts,
\begin{equation}
\mathbf{C}_{nl}(\bsi) = \frac{\bsi^{lin}}{\bsi} ~~,
\end{equation}
\noindent
where $\bsi^{lin}$ is the signal recorded by an ideal (linear) detector. 
Therefore, the values of the individual readouts of the pedestal and signal are given by
\begin{equation}
\bsi = \frac{\bsi^{lin}}{ \mathbf{C}_{nl}(\bsi)}  \ \ \ {\rm and} \ \ \ \bpi = \frac{\bpi^{lin}}{ \mathbf{C}_{nl}(\bpi)} ~~.
\end{equation}
\noindent
If the non-linearity of the detector is not accounted for, the estimated source flux $\bI$ can be substantially less than
the true source flux. Furthermore, as can be seen from Figure 1b, the values of the individual pedestals and signal readouts derived from Equations 5-9 will be underestimated.

The corrections $\mathbf{C}_{nl}$ for the non-linearity of optical and near-infrared arrays are often derived from flat-field data acquired expressly for this purpose.  A series of flat field images taken with gradually increasing exposure times are used to generate a curve of the recorded signal as a function of the exposure time, or equivalently estimated count rate as a function of recorded counts. (Of course, all systematic effects, such as those mentioned above, must be corrected for and removed before the non-linearity curve can be determined from the flat-field data.) The curve can then be fit with a continuous function (e.g., a polynomial), and the deviation of the function (at large count values for long exposures) from a straight line (fit to the count values recorded during short exposures) provides a measure of the non-linearity correction $C_{nl}(s_{k,l})$ as a function of the detected counts $s_{k,l}$ in pixel $k,l$.  For the SpeX array, for example, it was found that the non-linearity curve can be approximated extremely well at each pixel over the range $0$ to $\sim 8500$ DN (which represents $\sim 90$\% of the well depth and is close to saturation) by a rational function,

\begin{equation}
\mathbf{C}_{nl}(\mathbf{s}) \approx \frac{1}{1 + \mathbf{a_1} \cdot \mathbf{x} + \mathbf{a_2} \cdot \mathbf{x}^2 + \mathbf{a_3} \cdot \mathbf{x}^3} ~~,
\end{equation}

\noindent
where the $\mathbf{a_1}$,  $\mathbf{a_2}$, and $\mathbf{a_3}$ are constants,  $\mathbf{x} = \mathbf{s} - \mathbf{p}_{flat}$, and $\mathbf{p}_{flat}$ is the pedestal level estimated for the flat field images (using Equation \ref{eq:pest} below). This curve can then be used to correct the signals recorded by the detector (after removal of the bias, dark current, and any systematic effects) to the ``true'' values that would be recorded by a perfectly linear detector.

\subsection{Non-Linearity Corrections}

When attempting to apply a non-linearity correction to data acquired with CDS or MCDS techniques, it is important to keep in mind that the corrections must be made to {\it both} the signal and pedestal values separately, not to the net source counts, $\mathbf{S}_{net}$ or $\mathbf{S}_{tot}$.
However, because these readout procedures do not record separate signal and pedestal values (see \S2.1), we must estimate these {\it a posteriori} from $\mathbf{S}_{tot}$. Fortunately, this can be done relatively easily, with the definitions given in \S2.1, although in general an iterative procedure is required. 

We assume that the non-linearity corrections apply to the mean signal and pedestal values above the bias level, that is, the relative values above the bias rather than the absolute values. Furthermore, we assume the bias level remains constant during the integration and readouts, the dark current has been subtracted, and any other systematic effects have been corrected for before the non-linearity corrections are applied. In this case, first estimates of the pedestal and signal values can be calculated from 
\begin{eqnarray}
\mathbf{\bar{p}^{(1)}} & = & \frac{1}{n_c n_r}\sum_{j=1}^{n_c}\sum_{i=1}^{n_r} \mathbf{p}_{i,j} \\
                 & = &\frac{1}{n_c n_r}\sum_{j=1}^{n_c}\sum_{i=1}^{n_r} (i - \mathbf{f}) \frac{\mathbf{S}_{net,j}}{n_r}\frac{\delta t}{\Delta t}\\
                 & = & \frac{\bStot}{n_c n_r}\frac{\delta t}{\Delta t}\left [ \frac{n_r + 1}{2} - \mathbf{f} \right ] ~~.
%                 & = & \bI\delta t \left [ \frac{n_r + 1}{2} - \mathbf{f} \right ]
\label{eq:pest}
 \end{eqnarray}

\noindent
and
\begin{equation}
\mathbf{\bar{s}^{(1)}}  =  \frac{1}{n_c n_r}\bStot + \mathbf{\bar{p}^{(1)}}  ~~,
\label{eq:sest}
 \end{equation}
 \noindent
 respectively.

These values are then corrected for non-linearity using the $\mathbf{C}_{nl}$ curve

\begin{equation}
\mathbf{\bar{s}}^{(2)} = \mathbf{\bar{s}^{(1)}} \cdot \mathbf{C}_{nl}(\mathbf{\bar{s}^{(1)}}), \ \ \ 
\mathbf{\bar{p}}^{(2)} = \mathbf{\bar{p}^{(1)}} \cdot \mathbf{C}_{nl}(\mathbf{\bar{p}^{(1)}})
\end{equation}

\noindent
and a new estimate of the net source counts is computed from
\begin{equation}
\bStot^{(2)} = n_c n_r \mathbf{\bar{s}}^{(2)} - n_c n_r \mathbf{\bar{p}}^{(2)} ~~.
\end{equation}
\noindent 
It should be noted that the corrected values of $\mathbf{\bar{s}}$, $\mathbf{\bar{p}}$, and  $\bStot$ depend on the number of non-destructive reads, $n_r$ (see Eq.\ \ref{eq:pest}).

Using the corrected estimate of the total source counts $\bStot^{(2)}$, a new estimate of the mean pedestal level, $\mathbf{\bar{p}}^{(3)}$ can be computed from Equation \ref{eq:pest}. Adding this to 
 the {\em original} values of $\bStot$ using Equation \ref{eq:sest} yields a new estimate of the signal, $\mathbf{\bar{s}}^{(3)}$. The new estimates of $\mathbf{\bar{s}}$ and $\mathbf{\bar{p}}$ can then be corrected for non-linearity using the $\mathbf{C}_{nl}$ curve and then used to generate another estimate of true source counts $\bStot^{(3)}$. The procedure can be repeated as many times as necessary to achieve a desired convergence criterion.
 
For SpeX data we have found that convergence is rapidly achieved and that, 
unless the mean detected signal $\mathbf{\bar{s}}$ is close to the top of the well ($>8500$ DN, highly non-linear), the correction to the estimate of the true source counts is less than 1\% after three iterations;  therefore, $\bStot^{(4)} \approx \bStot^{lin}$.
An example of  both the necessity of making corrections for non-linearity and the effectiveness of the procedure outlined above can be seen in Figure 2, taken from the paper by  Cushing et al.\ (2003). This figure shows that spectra in adjacent spectral orders, reduced without incorporating the non-linearity corrections, do not match one another in either intensity level or spectral slope in the overlapping wavelength region. After the non-linearity corrections described above are made, however, the spectra agree in both intensity and slope to better than a few percent over the entire overlap region. 

\subsection{Variance Estimates}

Because the array is read out non-destructively in the CDS and MCDS (as well as the ``up-the-ramp'') techniques, the values recorded at a given pixel for each readout are correlated. Therefore, the variance $\bVI$ of the estimated source flux $\mathbf{I}$ cannot be estimated simply by the photon noise and the read noise  for the individual readouts added in quadrature. 
The covariance between any two reads must be included in the variance estimate \citep{garnett93}.

We start by considering the case where $n_c$=1.  The variance of $\bI$, $\bVI$, is given by the standard error propagation formula (Bevington 1969; Garnett \& Forrest 1993),

\begin{eqnarray}
\label{eq:vdef}
\bVI & = &  \sum_{i=1}^{n_r} \left [ \sigma^2_{\bsi} \left ( \frac{\partial \bI}{\partial \bsi} \right )^2 + \sigma^2 _{\bpi} \left ( \frac{\partial \bI}{\partial \bpi} \right )^2 \right ]  \\ \nonumber
        &   & + 2 \sum_{j=2}^{n_r} \sum_{i < j}  \left [ cov(\bsi,\bsj) \left ( \frac{\partial \bI}{\partial \bsi} \right ) \left ( \frac{\partial \bI}{\partial \bsj} \right ) + cov(\bpi,\bpj) \left ( \frac{\partial \bI}{\partial \bpi} \right ) \left ( \frac{\partial \bI}{\partial \bpj} \right ) \right ]  \\ \nonumber
        &   & +  2 \sum_{i=1}^{n_r} \sum_{j=1}^{n_r} \left [ cov(\bpi,\bsj) \left ( \frac{\partial \bI}{\partial \bpi} \right ) \left ( \frac{\partial \bI}{\partial \bsj} \right ) \right ] ~~,
\end{eqnarray}
\noindent
where $\sigma^2_{\mathbf{x}_i}$ is the variance of the $i$th readout value and $cov(\mathbf{x}_i,\mathbf{x}_j)$ is the covariance between the $i$th and $j$th read.  

The variance of the $i$th readout value is given by the sum of the square of the photon noise and the square of the read noise. The read noise, $\boldsymbol{\sigma}_{read}$, is usually specified in terms of electrons. If the array has a gain $\bg$ (with units of electrons per data number, e$^-$ DN$^{-1}$),
conversion to DN gives a variance associated with the read noise of $\boldsymbol{\sigma}_{read}^2/\bg^2$. For Poisson noise, the variance of each readout value, expressed in DN$^2$, is given by
\begin{equation}
\sigma^2_{\bsi} = \frac{\bsi}{\bg} \ \ \ {\rm and} \ \ \ \sigma^2_{\bpi} = \frac{\bpi}{\bg} ~~ .
\end{equation}

The covariance between the $i$th and $j$th readout is given by the square of the photon noise of the $i$th read for $j>i$ \citep{garnett93}. This can be easily shown as follows. For any two reads $i,j$, with $j>i$, the associated readout values are $\mathbf{x}_i$ and $\mathbf{x}_j$, which are related by 
\begin{equation}
\mathbf{x}_j = \mathbf{x}_i + \Delta_{j-i}
\end{equation}
where $\Delta_{j-i}$ is the difference in counts between the two reads. Then
\begin{eqnarray}
cov(\mathbf{x}_i,\mathbf{x}_j) & = & \langle(\mathbf{x}_i - \langle \mathbf{x}_i \rangle)\cdot (\mathbf{x}_j - \langle \mathbf{x}_j\rangle)\rangle \\ \nonumber
%                               & = & \langle \mathbf{x}_i \cdot \mathbf{y}_j\rangle - \langle\mathbf{x}_i\rangle \cdot \langle\mathbf{y}_j\rangle \\
 %                              & = & \langle \mathbf{x}_i \cdot (\mathbf{x}_i + \Delta)\rangle - \langle\mathbf{x}_i\rangle \cdot \langle(\mathbf{x}_i + \Delta)\rangle \\ \nonumber
                               & = & \langle \mathbf{x}_i^2\rangle - \langle \mathbf{x}_i\rangle^2 + 
 \langle \mathbf{x}_i \cdot \Delta_{j-i}\rangle - \langle \mathbf{x}_i\rangle  \cdot \langle \Delta_{j-i}\rangle \\ \nonumber
                               & = & cov(\mathbf{x}_i,\mathbf{x}_i) + cov(\mathbf{x}_i,\Delta_{j-i}) \\ 
                               & = & \sigma_{\mathbf{x}_i}^2
\end{eqnarray}
where the angle brackets $\langle ~~ \rangle$ denote the average.
\noindent
Therefore, we have
\begin{equation}
cov(\mathbf{s}_i,\mathbf{s}_j) = \sigma_{\mathbf{s}_i} \ \ \ {\rm and} \ \ \ cov(\mathbf{p}_i,\mathbf{p}_j) = \sigma_{\mathbf{p}_i}    {\rm for ~~ j > i}
\end{equation}
and
\begin{equation}
cov(\mathbf{p}_i,\mathbf{s}_j) = \sigma_{\mathbf{p}_i} ~~~.
\end{equation}

Upon substitution, Equation \ref{eq:vdef} becomes,
\begin{eqnarray}
\bVI & = & \frac{1}{n_r^2 \Delta t^2} \left \{ \sum_{i=1}^{n_r} \left ( \frac{\bsi}{\bg} + \frac{\boldsymbol{\sigma}^2_{read}}{\bg^2} + \frac{\bpi}{\bg} + \frac{\boldsymbol{\sigma}^2_{read}}{\bg^2} \right ) + 2 \sum_{j=2}^{n_r} \sum_{i=1}^{j-1} \left( \frac{\bsi}{\bg} + \frac{\bpi}{\bg} \right) -2 \sum_{i=1}^{n_r} \sum_{j=1}^{n_r} \frac{\bpi}{\bg} \right \} \\
        & = & \frac{1}{\bg n_r^2 \Delta t^2} \left \{ \sum_{i=1}^{n_r} (\bsi + \bpi) + 2 \sum_{i=1}^{n_r} (n_r - i) (\bsi + \bpi)  - 2n_r \sum_{i=1}^{n_r} \bpi + \frac{2n_r}{\bg}\boldsymbol{\sigma}^2_{read} \right \} \\ 
        & = & \frac{1}{\bg n_r^2 \Delta t^2} \left \{ \sum_{i=1}^{n_r}(\bsi + \bpi) + 2n_r\sum_{i=1}^{n_r}\bsi - 2\sum_{i=1}^{n_r} i (\bsi + \bpi) + \frac{2n_r}{\bg}\boldsymbol{\sigma}^2_{read} \right \} ~~.
\end{eqnarray}

We now assume that the array has a perfectly linear response, such that $\bsi=\bsi^{lin}$, 
$\bpi=\bpi^{lin}$, and $\bSnet=\bSnet^{lin}$. If the array is not linear,
the signal and pedestal images must be first corrected for non-linearity as outlined above.
Explicitly accounting for non-linearity would introduce factors of $\mathbf{C}_{nl}(\bsi)$ and $\mathbf{C}_{nl}(\bpi)$ in the equations given below, which would result in an increase in
the variance of any given pixel value beyond that expected for an intrinsically linear array and would also
prevent us from deriving completely general equations for the variance in terms of recorded quantities. 
However, as we will demonstrate below (\S4), this increase can be easily accounted for by modifying the values of the gain and read noise from their ``intrinsic'' values. 

Substituting the definitions of $\bsi$ and $\bpi$ for a linear, stable (constant $\mathbf{b}$) detector from Equations 5 and 7 into Equation 36, we have,
\begin{eqnarray}
\label{eq:vintermed}
\bVI & = & \frac{1}{\bg n_r^2 \Delta t^2} \left \{\sum_{i=1}^{n_r} \left ( \frac{\bSnet}{n_r} + (i - \mathbf{f})\frac{2\bSnet}{n_r}\frac{\delta t}{\Delta t} \right) + 2n_r\sum_{i=1}^{n_r} \left (\frac{\bSnet}{n_r} + (i - \mathbf{f})\frac{\bSnet}{n_r}\frac{\delta t}{\Delta t} \right )  \right. \nonumber \\
        &   & - \left. 2\sum_{i=1}^{n_r} i \left (\frac{\bSnet}{n_r} + (i - \mathbf{f})\frac{2\bSnet}{n_r}\frac{\delta t}{\Delta t} \right ) + \frac{2n_r}{\bg}\boldsymbol{\sigma}^2_{read} \right \}~~.
%        & = & \frac{1}{n_r^2 \Delta t^2} \left \{ \frac{1}{\bg}\bSnet + \frac{1}{\bg}\bSnet \frac{\delta t}{\Delta t} (n_r+1) + \frac{2n_r}{\bg}\mathbf{b} + \frac{2n_r}{\bg}\bSnet + \frac{1}{\bg}\bSnet \frac{\delta t}{\Delta t} n_r(n_r + 1)  \right. \\ \nonumber
%        &   & \left. + \frac{2n_r^2}{\bg}\mathbf{b} - \frac{1}{\bg}\bSnet(n_r + 1) - \frac{2}{3\bg}\bSnet \frac{\delta t}{\Delta t} (n_r + 1)(2n_r+1) - \frac{2n_r(n_r+1)}{\bg}\mathbf{b}+ \frac{2n_r}{\bg^2}\boldsymbol{\sigma}^2_{read} \right \} ~~,
\end{eqnarray}
\noindent
Using the relations
\begin{equation}
\sum_{i=1}^{n} i = \frac{1}{2}n(n+1), \ \ \ \sum_{i=1}^{n} i^2 = \frac{1}{6}n(n+1)(2n+1) ~~,
\end{equation}
\noindent
we can simplify Equation \ref{eq:vintermed} as follows,
\begin{eqnarray}
\bVI & = & 
%\frac{1}{n_r^2 \Delta t^2} \left \{ \frac{n_r}{\bg}\bSnet - \frac{\bg}{3}\bSnet \fdt (n_r+1) \left [n_r-1 \right ] + \frac{2n_r}{\bg^2}\boldsymbol{\sigma}^2_{read} \right \} \\
 \frac{\bSnet}{\bg n_r \Delta t^2} \left [ 1 - \frac{1} {3} \fdt \frac{(n_r^2-1)}{n_r} \right ] + \frac{2\boldsymbol{\sigma}^2_{read}}{\bg^2n_r \Delta t^2}  \\
        & = & \frac{\bI}{\bg \Delta t} \left [ 1 - \frac{1} {3} \fdt \frac{(n_r^2-1)}{n_r}\right ] + \frac{2\boldsymbol{\sigma}^2_{read}}{\bg^2n_r \Delta t^2} ~~.
\end{eqnarray}
\noindent
Note that the factor $\mathbf{f}$ drops out of the variance estimate.

We now consider the case of $n_c$ co-additions.  The variance then becomes,
\begin{eqnarray}
\bVI & = & \frac{1}{n_c^2}\sum_{i=1}^{n_c}\mathbf{V}_{I_{i}}  \\
     & = & \frac{1}{n_c^2}\sum_{i=1}^{n_c} \left ( \frac{\mathbf{S}_{net,i}}{\bg n_r \Delta t^2} \left [1 - \frac{1}{3} \fdt \frac{(n_r^2-1)}{n_r} \right ] + \frac{2\sigma^2_{read}}{\bg^2n_r \Delta t^2} \right ) \\
     & = & \frac{\bStot}{\bg n_r n_c^2\Delta t^2} \left [ 1 - \frac{1}{3} \fdt \frac{(n_r^2-1)}{n_r} \right ] + \frac{2\boldsymbol{\sigma}^2_{read}}{\bg^2n_r n_c\Delta t^2} \\
     & = & \frac{\bI}{\bg n_c\Delta t} \left [ 1 - \frac{1}{3} \fdt \frac{(n_r^2-1)}{n_r} \right ] + \frac{2\boldsymbol{\sigma}^2_{read}}{\bg^2n_r n_c\Delta t^2}~~.
     \label{eq:var}
\end{eqnarray}

Finally, let us define the {\em effective read noise} $\sigma_{read, eff}$ as
\begin{equation}
\boldsymbol{\sigma}_{read, eff} = \frac{\sqrt{2}\boldsymbol{\sigma}_{read}}{\sqrt{n_r n_c}}
\label{eq:effrn}
\end{equation}
which has the units of electrons. The variance on $\bI$ is then given by
\begin{equation}
\bVI = \frac{\bI}{\bg n_c\Delta t} \left [ 1 - \frac{1}{3} \fdt \frac{(n_r^2-1)}{n_r} \right ] + \frac{\boldsymbol{\sigma}^2_{read, eff}}{\bg^2 \Delta t^2} ~~~.
\end{equation}

Systematic effects produce additional terms in the equations for the variances. If these effects are
present, but not accounted for (as in our analysis), they will manifest themselves as an increased (i.e., larger than intrinsic) or perhaps variable ``read noise'' (see \S 4).

\section{Continuous Sampling Technique}
\subsection{Ideal Detector}

The notation given in \S 2 can also be used to estimate the pixel variances of an ideal detector read out with the  ``continuous sampling'' (also known as the line-fitting or ``up-the-ramp") technique \citep{chapman90, finger00}. In this technique, the array is read out repeatedly during the entire exposure. The count values from the readouts are then fit with a straight line and the slope gives the estimated source flux, $\mathbf I$ (in DN s$^{-1}$). For $n_r$ reads, each of duration $\delta t$, during an exposure of $\Delta t$, the slope can be written as follows \citep{bevington69}:
\begin{equation}
\mathbf{I} = \frac{n_r \sum_{i=1}^{n_r} \bti \bsi - \sum_{i=1}^{n_r} \bti \sum_{i=1}^{n_r} \bsi}{n_r\sum_{i=1}^{n_r}\bti^2 - (\sum_{i=1}^{n_r} \bti)^2}
\end{equation}
where $\bti$ is the time of the $i$th readout, and $\bsi$ is the count value recorded during this readout. This equation assumes that all readouts are given equal weight in the fitting process. If the readouts are equally spaced in time, $\bti = (i - \mathbf{f})\delta t$,  $\Delta t = (n_r -1) \delta t$, and the above equation can be simplified to yield,
\begin{equation}
\mathbf{I} = \frac{\sum_{i=1}^{n_r} \bsi (i - \frac{n_r + 1}{2})}{\alpha}
\label{eq:cslope}
\end{equation}
where $\alpha = n_r (n_r + 1) \Delta t/12$ . (We refer the reader to the report by Sparks (1998) for the case where the readouts are not equally spaced in time.) For a perfectly linear detector exposed to a source with constant flux, the values recorded at the end of each individual read are given by
\begin{equation}
\bsi = \bI t_i  = (i - \mathbf{f})\bI \delta t  ~~.
\end{equation}

\subsection{Non-linearity corrections}
Non-linearity corrections are fairly straight-forward to implement with this readout technique. 
Again, we assume that the bias level is constant, the bias and dark current have been subtracted, and systematic effects have been accounted for before the non-linearity corrections are applied. 
Equation 49 yields the first estimate of the signal values, $\bsi^{(1)}$ for each read.
The non-linearity corrected signal values can then be estimated in a manner similar to that for Fowler sampling,

\begin{equation}
\bsi^{(2)} = \bsi^{(1)} \cdot \mathbf{C}_{nl}(\bsi^{(1)}) ~~.
\end{equation}
Once all the signal readout values for all $i$ reads have been corrected, the corrected
slope can be re-determined from Equation \ref{eq:cslope}. The procedure can then be
repeated to yield more accurate values of $\bsi$ and $\bI$, until a suitable convergence criterion is reached.

\subsection{Variances}
From Equation (\ref{eq:cslope}) the variance in the slope recorded by a linear array  can be calculated in a manner similar to that used to determine the uncertainties in MCDS:
\begin{eqnarray}
\bVI & = & \sum_{i=1}^{n_r}\boldsymbol{\sigma}_{\bsi}^2 \left (\frac{\partial \mathbf I}{\partial \bsi}\right )^2  +  2 \sum_{j=2}^{n_r}\sum_{i < j}cov(\bsi,\mathbf {s_j})\left (\frac{\partial \mathbf I}{\partial \bsi}\right )\left (\frac{\partial \mathbf I}{\partial \mathbf s_j}\right ) \\
         & = & \sum_{i=1}^{n_r}\left (\frac{\bsi}{\mathbf{g}} + \frac{\boldsymbol{\sigma}_{read}^2}{\mathbf{g}^2}\right )\left[\frac{i - \frac{1}{2}(n_r + 1)}{\alpha}\right]^2 + \nonumber \\
         &     & 2 \sum_{j=2}^{n_r}\sum_{i =1}^{j-1}\frac{\bsi}{\mathbf{g}} \left [\frac{i - \frac{1}{2}(n_r + 1)}{\alpha}\right ]\left [\frac{j - \frac{1}{2}(n_r + 1)}{\alpha}\right]~,
\end{eqnarray}
\noindent
Using the definition of $\bsi$ above (Eq. 49), along with relations given in Equation 38 and the additional definitions

\begin{equation}
\sum_{i=1}^{n} i^3 = \frac{n^2(n+1)^2}{4} ,  \ \ \ \ \sum_{i=1}^{n} i^4 = \frac{n}{30}(n+1)(2n + 1)(3n^2 + 3n - 1)~~~,
\end{equation}

\noindent we find 
\begin{eqnarray}
\bVI & = & \frac{\mathbf{I}\delta t}{\mathbf{g}}\frac{n_r (n_r^4 - 1)}{120 \alpha^2} + \frac{\boldsymbol{\sigma}_{read}^2}{\mathbf{g}^2} \frac{n_r(n_r^2 - 1)}{12 \alpha^2} \\
        & = & \frac{6}{5}\frac{\mathbf{I}}{\mathbf{g}n_r \Delta t}\left(\frac{n_r^2 + 1}{n_r + 1}\right) +  \frac{12 \boldsymbol{\sigma}_{read}^2}{\mathbf{g}^2 n_r \Delta t^2}\left(\frac{n_r - 1}{n_r + 1}\right) \\
        & = & \frac{6}{5}\frac{\mathbf{I}}{\mathbf{g}n_r \Delta t}\left(\frac{n_r^2 + 1}{n_r + 1}\right) +  \frac{6 \boldsymbol{\sigma}_{read, eff}^2}{\mathbf{g}^2 \Delta t^2}\left(\frac{n_r - 1}{n_r + 1}\right) ~~,
\end{eqnarray}
\noindent
where we have used the definition of the effective read noise given earlier (Eq. 45).
These equations are identical to that given by Garnett \& Forrest (1993) once the different
definitions of the exposure time are taken into account. However, they are substantially
different from that derived by Chapman et al.\ (1990), who did not account for the 
correlation of the individual readouts in their analysis.

\section{Verification and Application}

To verify that the relation between observed count rate and variance predicted by Equation \ref{eq:var} for the MCDS technique yields realistic estimates of the variance in any given pixel, we obtained a series of flat field exposures with the SpeX instrument (whose array is read out using Fowler sampling) in the SXD mode \citep{rayner03}.  (Unfortunately, because we lack easy access to an array which is read out with the continuous readout technique, we were unable to obtain similar data to verify the procedures and equations given in \S 3.)
Details about the readout procedures and electronics for SpeX can be found in \citet{rayner03} and references therein. Because SpeX is a cross-dispersed spectrograph, the flat fields exhibit a large range in observed count values, which makes them ideal for our purposes. Two series of exposures were taken with $\Delta t = 17$ s,  $\delta t = 0.51$ s, and successive $n_r$ values of $1, 4, 8, 16, 32, 1$ and $32, 16, 8, 4, 1, 32$. 
In all cases, $n_c = 1$. For each combination of $\Delta t$ and $n_r$, nineteen flat field exposures were obtained, from which a mean count rate value and a standard deviation were computed at each pixel. Non-linearity corrections were then applied to the individual frames and the means and standard deviations were  re-computed at each pixel.
 
For plotting purposes only, we computed the mean values and standard deviations of the variances in bins of the observed count values that were 50 DN wide. In Figure 3a, the mean variances are plotted versus the mean count values for one of the $n_r = 1$ sets of flats. The results from the original data are shown as well as  those obtained after applying linearity corrections. The same plot for one of the $n_r=32$ sets of flats is shown in Figure 3b. As can be seen from these figures, a linear trend of the variances of the non-linearity-corrected data with the observed counts, like that predicted by Equation \ref{eq:var}, provides a very good representation of the results.
 
Equation \ref{eq:var} was then fit to the actual (i.e., un-binned) flat field data, with a least squares procedure in which all data points were equally weighted, to determine the  array-averaged gain $g$ and read noise $\sigma_{read}$ for each set of flats (i.e., for each set of $\Delta t$ and $n_r$ values). 
We included only those pixels with values between $0 < S_{tot}/n_r < 4000$ DN in our analysis. Given the exposure parameters for the flat fields, this range of values represents most of the usable well for the SpeX array. Because the non-linearity corrections depend on $n_r$, the corrections for the $n_r = 1$ data sets are less than $\sim 5$\%, while the non-linearity corrections for the $n_r = 32$ data sets are less than about $15$\%, for $S_{tot}/n_r < 4000$ DN. (These estimates have been derived from the SpeX non-linearity curve and simulations we have performed with SpeX flat field images.) SpeX documentation strongly advises users to keep values of $S_{tot}/n_r$  below 4000 for all observations. The best fits are shown as solid lines on Figures 3a and 3b. Since most of the data points have values $< 1500$ DN, the fits are heavily biased to these lower values. (The distribution of $S_{tot}/n_r$ values in the flat fields are shown at the top of the two plots.) 

The gain and readnoise values resulting from the fits are shown in Figure 4, plotted as a function of the number of reads. The values for the gain (Fig. 4a) were found to be reasonably consistent with one another, with a mean of 12.1 e$^-$ DN$^{-1}$, although there is some indication of slight decrease in
$g$ as $n_r$ increases. We note that the smallest gain values were derived from the first images in both series of exposures, which were taken after the instrument and detector had been sitting idle for a long period. As seen in Figure 4b, however, there is a clear correlation between $\sigma_{read}$ and $n_r$. The variation in $\sigma_{read}$ with $n_r$ is well described by a power law with an index of 0.16,

\begin{equation}
\sigma_{read} \approx 36  \cdot n_r^{0.16} ~~~.
\end{equation}

\noindent
This relation implies that the effective read noise ($\sigma_{read, eff}$, Equation \ref{eq:effrn}) does not decrease as the square root of the number of reads, as might be expected, but rather decreases at a much slower rate, as $n_r^{-0.34}$. This is demonstrated in Figure 5 where we plot the effective read noise determined from the flat field data and the best fitting power law. 

An alternative description of the variation of the effective read noise with the number of reads is 
the following,
\begin{equation}
\label{eq:effrnalt}
\sigma_{read, eff} = A \cdot n_r^{-0.5} + {\rm Constant}~~~,
\end{equation}
where $A$ is a constant proportional to the read noise $\sigma_{read}$.
In this formulation, the read noise $\sigma_{read}$ does have a constant value and the effective read noise does decrease as $n_r^{-0.5}$, but there is an additional constant noise source that results in an additive offset. A fit of this form to the SpeX data is also shown in Figure 5, where it can be seen that it also provides an acceptable representation of the data.

The variation of the effective read noise with $n_r$ determined from the SpeX data appears remarkably similar to that presented in Figure 1 of Garnett \& Forrest (1993). Although these authors claim that the effective read noise decreases as $n_r^{-0.5}$, close inspection of their figure reveals that a power law with $n_r^{-0.5}$ cannot reproduce the data points without an additive constant offset (Equation \ref{eq:effrnalt}).  A power law with a smaller index of about 0.4 would probably also fit their data well.  

As mentioned in \S 2, variation in the read noise with $n_r$ is a sign that systematic effects in the behavior of the array or the readout procedure are present but not accounted for in our analysis. Although there are several possible candidates for such an overlooked systematic effect (see \S2), the exact cause for the observed variation of the read noise with the number of reads is uncertain at present. The variation is probably not a result of the non-linearity corrections, as the fits are biased to a range of data values for which the non-linearity corrections are negligible. Furthermore, variations in the read noise with the number of reads have been observed before by others. For example, Finger et al.\ (2000) have suggested that the deviation from $n_r^{-0.5}$ in the effective read noise measurements of the Aladdin array installed in the ISAAC instrument at the VLT  is due to additional Poisson noise from multiplexer glow and dark current. As the dark current is implicitly included in our formulation for the variance, increased dark current in the SpeX array is probably not the source of the additional noise. Inspection of dark frames generated by SpeX reveals no evidence of amplifier glow. Furthermore, it is straightforward to include an additional source term in  Equations (5-9) to represent amplifier glow, which is present only during the readouts. This term leads to an additional noise term in the expression for the variance that increases as $n_r$, which is significantly larger than the variation we observe. 

However, additional noise might result from the slight heating of the array due to the numerous reads. In order
to reduce residual image effects the SpeX array is read out at 1 Hz when not integrating \citep{rayner03}. The act of reading out the array dissipates ~20mW. An exposure consists of an initial burst (2 Hz) of reads of the pedestal (heating), followed by a period of almost no power dissipation
during the actual integration (cooling), terminated by a burst (2 Hz) of reads
of the signal (heating). Although the array mount is controlled to $30.00 \pm 0.01$K, power
is dissipated locally in the array, and so the variable heating effects of the readout might possibly cause much larger temperature differences. Changes in temperature (or the resulting instabilities) 
as a function of the number of reads would lead to a variation in the dark current rate and/or the bias level during the reads and therefore might explain the observed variation in read noise (see e.g., B\"oker et al.\ 2001). This may well be a common phenomenon in many readout techniques for near-infrared arrays.

Because we have not explicitly included these systematic effects, or the corrections for non-linearity,
in our equations, we should point out that the gain and read noise derived from our fits are not necessarily the ``true", or intrinsic, values for the array. The non-linearity corrections of the individual images, for example,  increase the variance of any given pixel value above that expected for the same pixel value recorded on an intrinsically linear array. Therefore, the estimated gain and read noise values determined from the fits will generally be smaller and larger, respectively, than the intrinsic values for the array.
As far as data reduction is concerned, however, this does not pose a problem, as long as all images subject to the same systematic effects are reduced with the same parameter values. For example, the {\it Spextool} software first corrects each individual image for non-linearity before generating a variance image or proceeding with any reduction steps (see Cushing et al.\ 2003). The increased variances due to the non-linearity corrections are then correctly accounted for by adopting the gain and read noise values derived from our linear fits. Hence, these gain and read noise values are necessary for consistency in the data reduction routines.

\section{Summary}

We have derived fairly simple equations for the estimated variance as a function of the
observed count rate at each pixel in a near-infrared array for correlated double sampling,
multiple correlated double sampling, and equally-spaced continuous sampling read out techniques. We
also give a general prescription for implementing corrections for non-linearity at each
pixel. We have applied these equations to sets of flats obtained with the SpeX instrument
at the NASA Infrared Telescope Facility and find that they provide a good representation to the 
non-linearity-corrected data. We also find that the read noise of the SpeX
array varies with the number of reads as $n_r^{0.16}$, perhaps due to heating of the 
array as a result of multiple reads. Such heating may be generic to near-infrared arrays that employ (M)CDS as the readout technique.

\acknowledgments

 W.\ Vacca thanks Alan Tokunaga and James Graham for support and for reading an earlier draft of this paper. M.\ Cushing acknowledges financial support from the NASA Infrared Telescope facility.  We thank J.\ Leong and  P.\ Onaka for acquiring some of the SpeX flat fields used in our analysis. We also thank P.\ Onaka for discussions regarding the operation of near-infrared array electronics and readout techniques.

\clearpage

\begin{figure}
\psfig{file=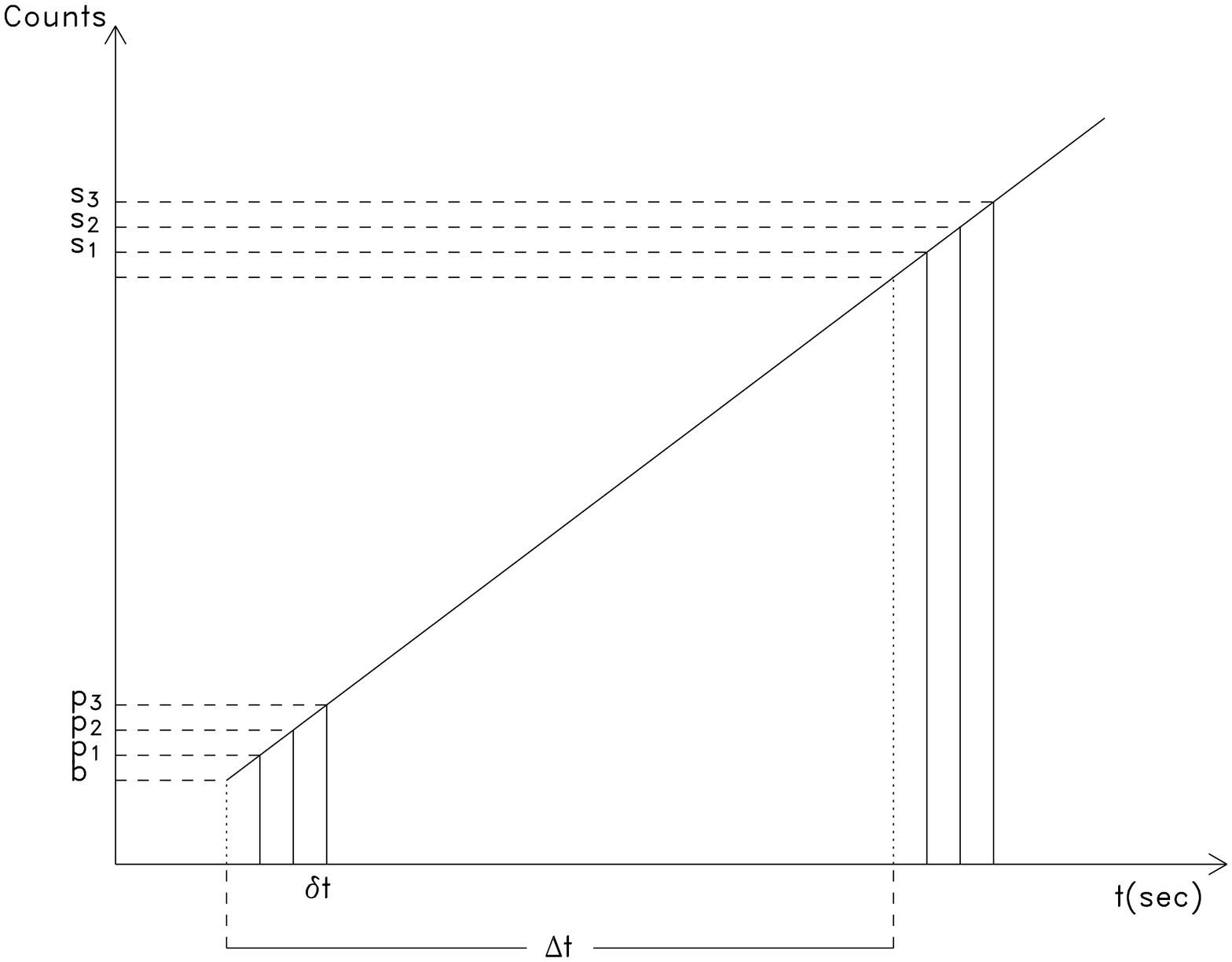,angle=90,height=5.5in}
\caption{Schematic of the signal recorded by an integrating node as a function of time on ({\em a}) a perfectly linear detector and ({\em b}) a real (i.e., non-linear) detector  employing the MCDS read out technique. The two dotted vertical lines represent the starting time of the first read of the pedestal and the start of the first read of the signal. Each read is assumed to last $\delta t$ sec. Each solid vertical line represents the end of a read. The exposure time is $\Delta t$ sec.} 
\end{figure}

\begin{figure}
\psfig{file=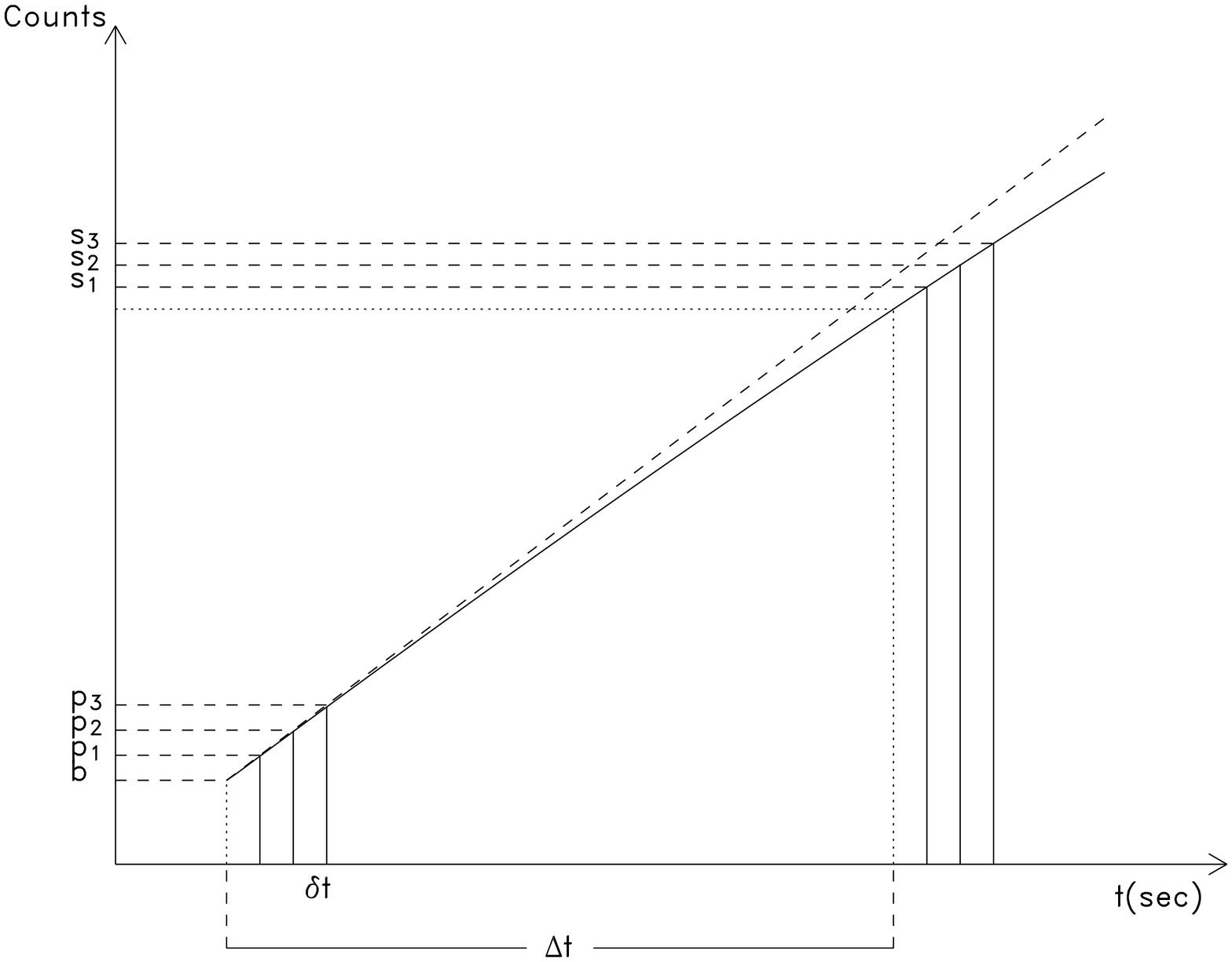,angle=90,height=5.5in}
\end{figure}

\begin{figure}
\psfig{file=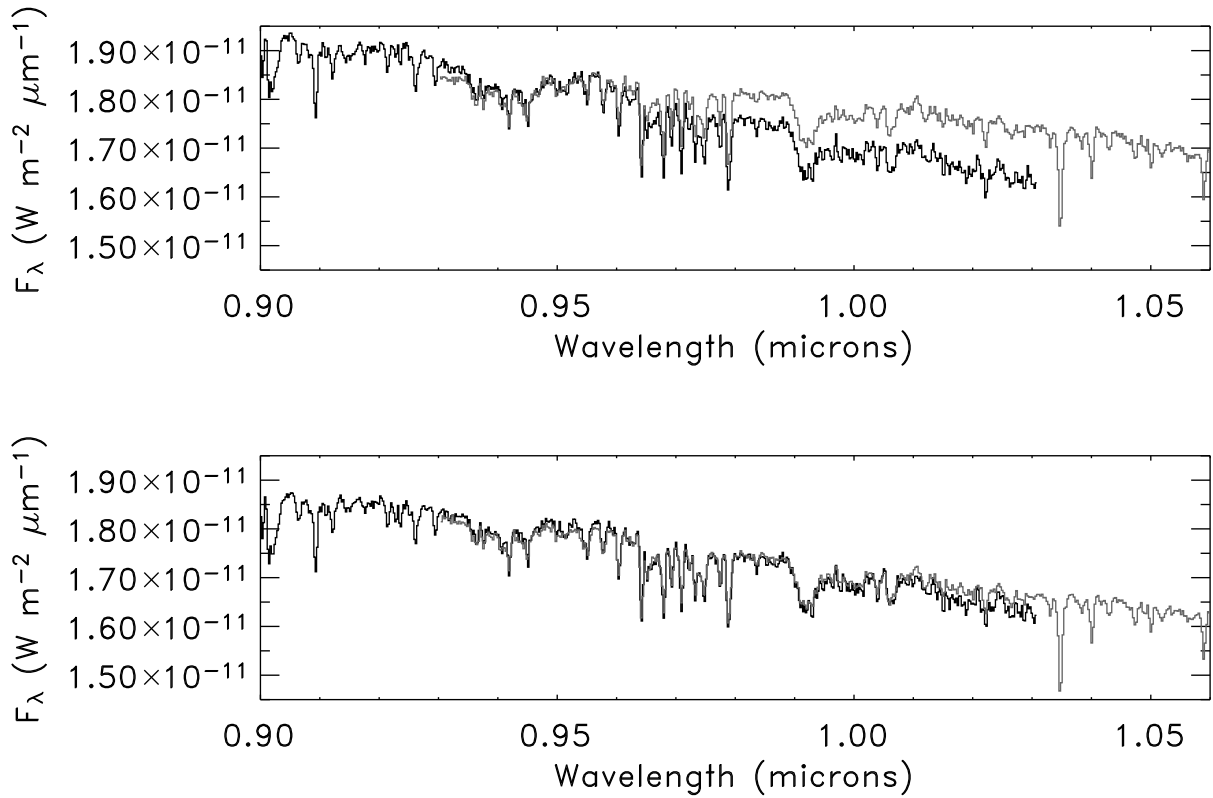,angle=0,height=4.5in}
\caption{Spectra of Gl 846 from orders 6 and 7 of SpeX in the SXD mode (see Rayner et al.\ 2003). 
The spectra were reduced with the {\it Spextool} reduction package (Cushing et al.\ 2003). The spectra shown in the upper panel were reduced without applying the non-linearity corrections discussed in the text; the corrections were incorporated in the reduction of the spectra shown in the lower panel. Incorporation of the non-linearity corrections removes to a large degree the mismatch in both the flux levels and the slopes of the spectra in the overlap regions. }
\end{figure}

\begin{figure}
\psfig{file=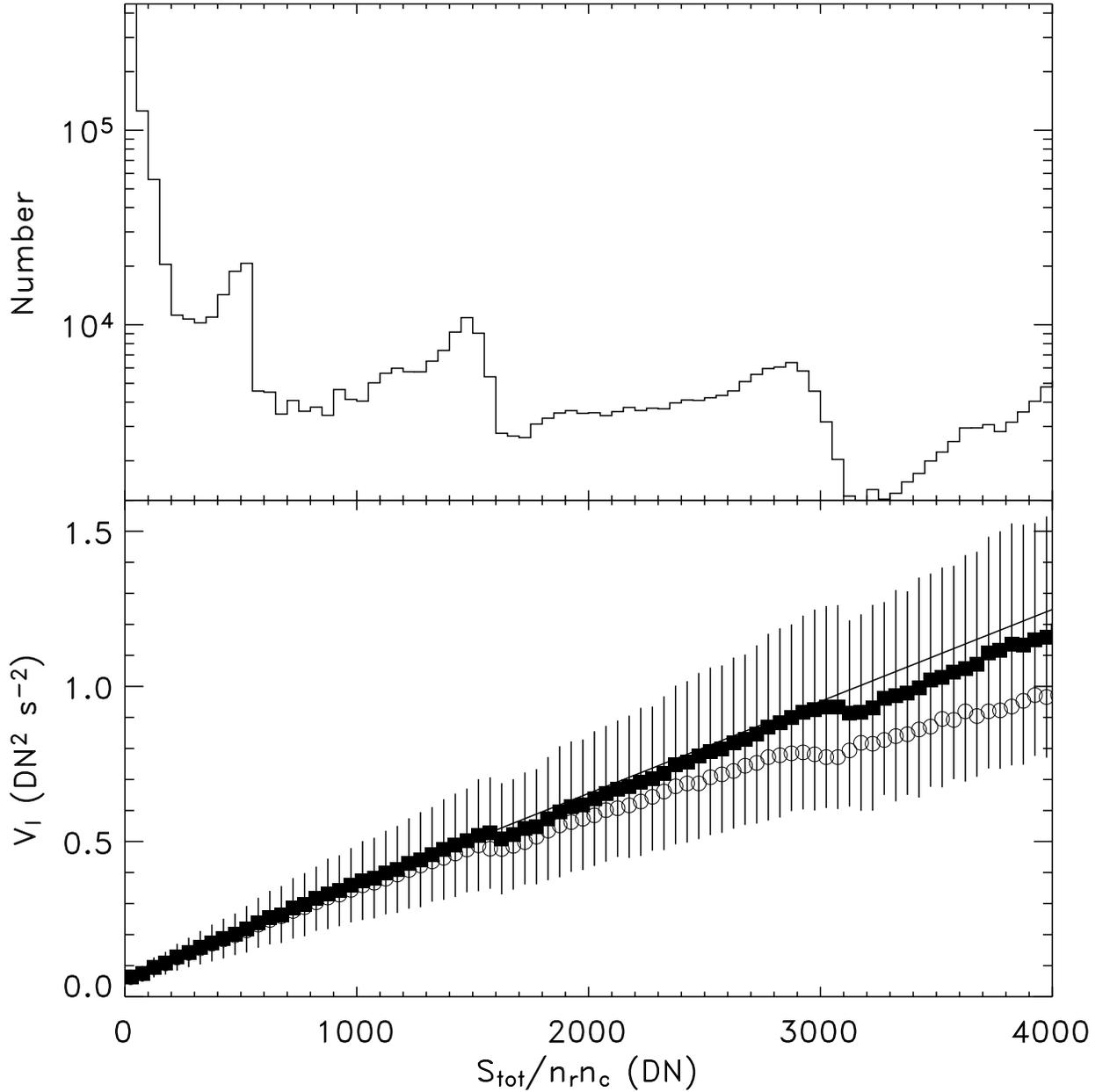,angle=0,height=7in}
\caption{Measured mean variances and standard deviations as a function of count value for two sets of flat field data, ({\em a}) $n_r = 1$  and ({\em b}) $n_r =32$. The values have been computed in bins of 50 DN wide. The open points are the original values, while the solid points represent the values measured after making non-linearity corrections to the individual frames (see text). The solid line represents a fit of Eq. 47 to the
non-linearity-corrected data points. Since most of the data have count values below 1500 DN, the fit is heavily biased to match the variances in this range. The histogram of data values is shown at the top of both plots. }
\end{figure}

\begin{figure}
\psfig{file=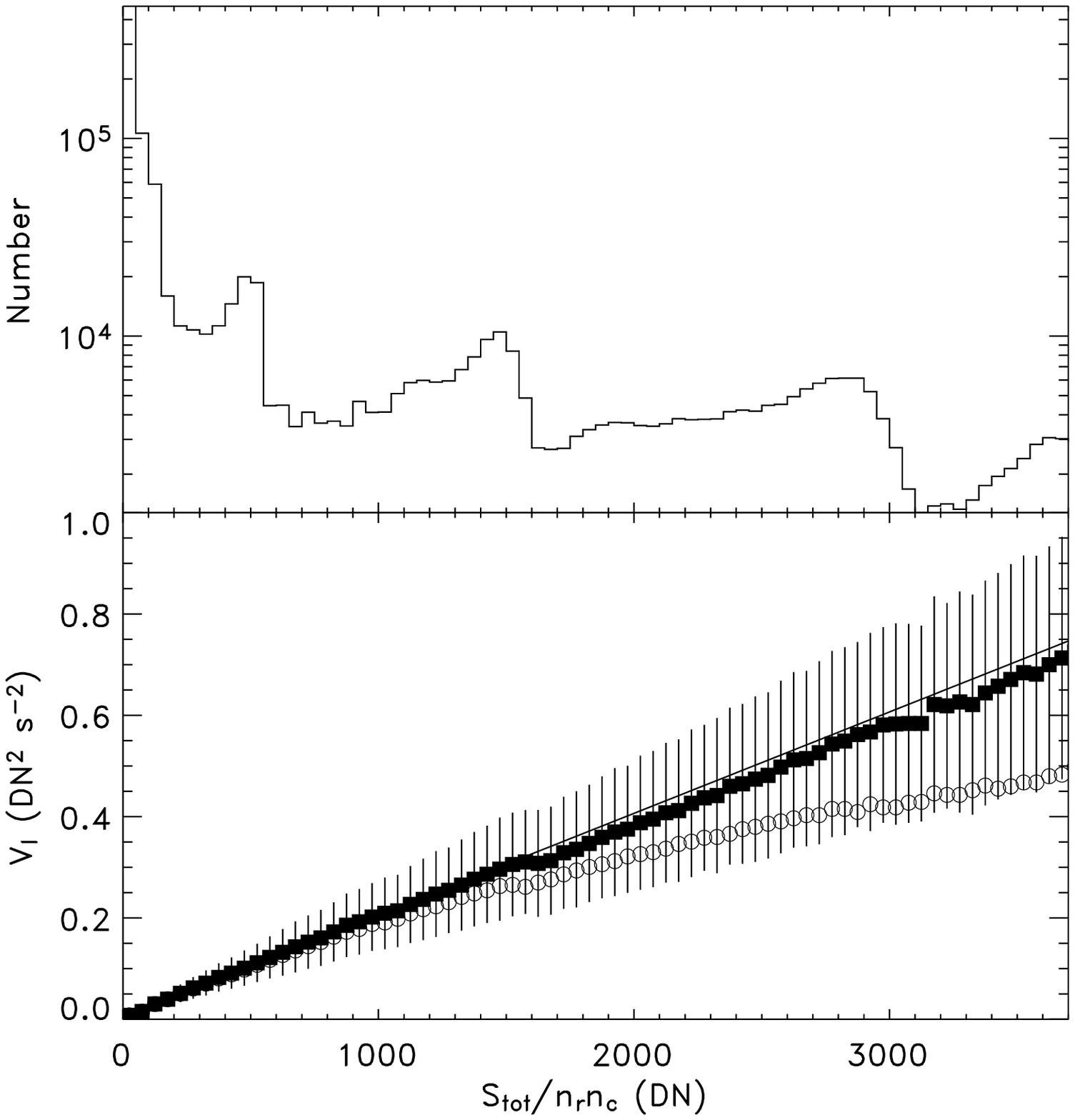,angle=0,height=7in}
\end{figure}

\begin{figure}
\psfig{file=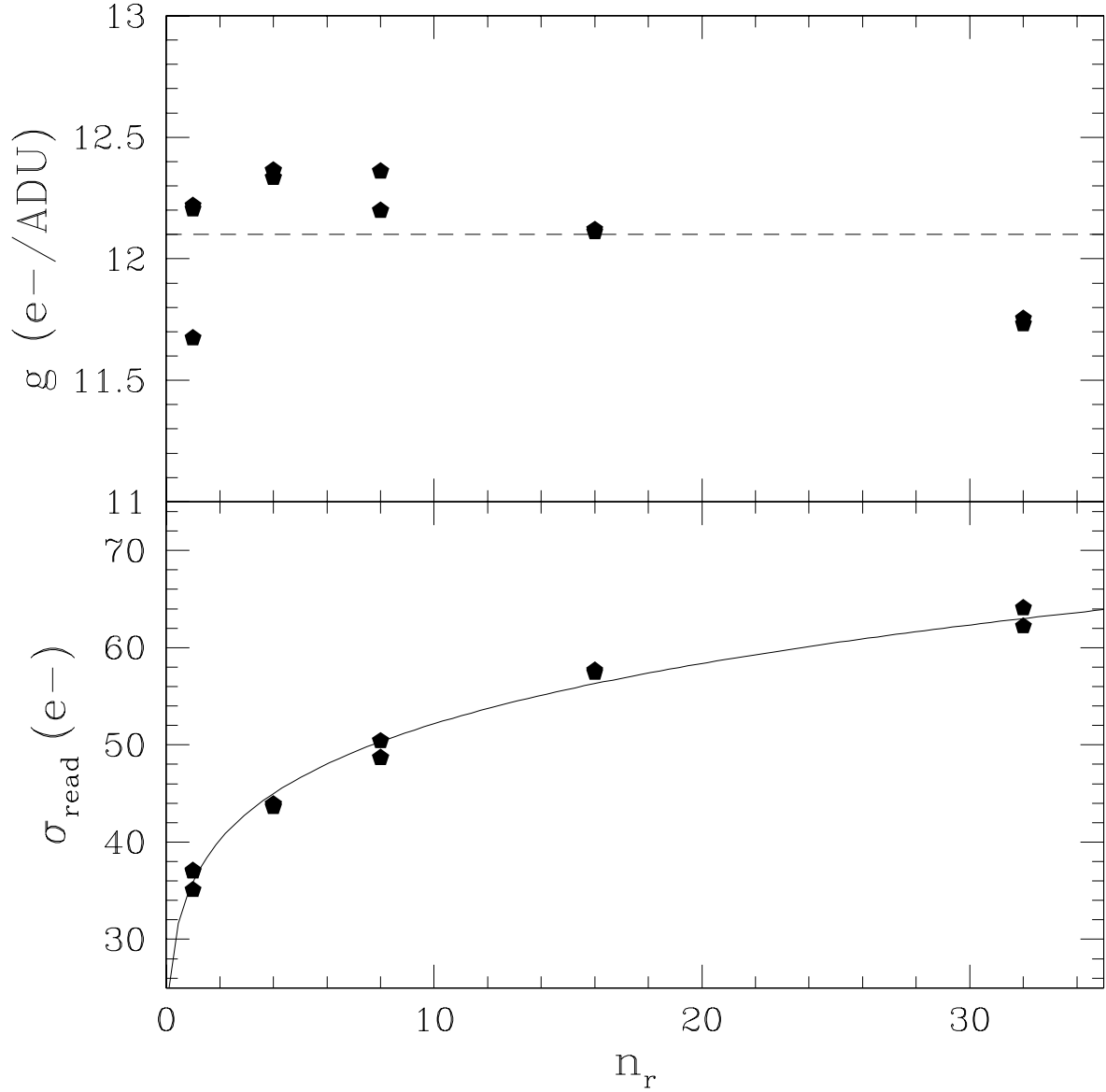,angle=0,height=6.5in}
\caption{({\em top}) Gain (in electrons/DN) and ({\em bottom}) read noise (in electrons) measured for the 
SpeX array as a function of the number of reads. The dashed line is the estimated mean value of the gain (12.1); the solid line is a fit of a power law to the read noise as a function of the number of reads.
The power law index is 0.16.}
\end{figure}

\begin{figure}
\psfig{file=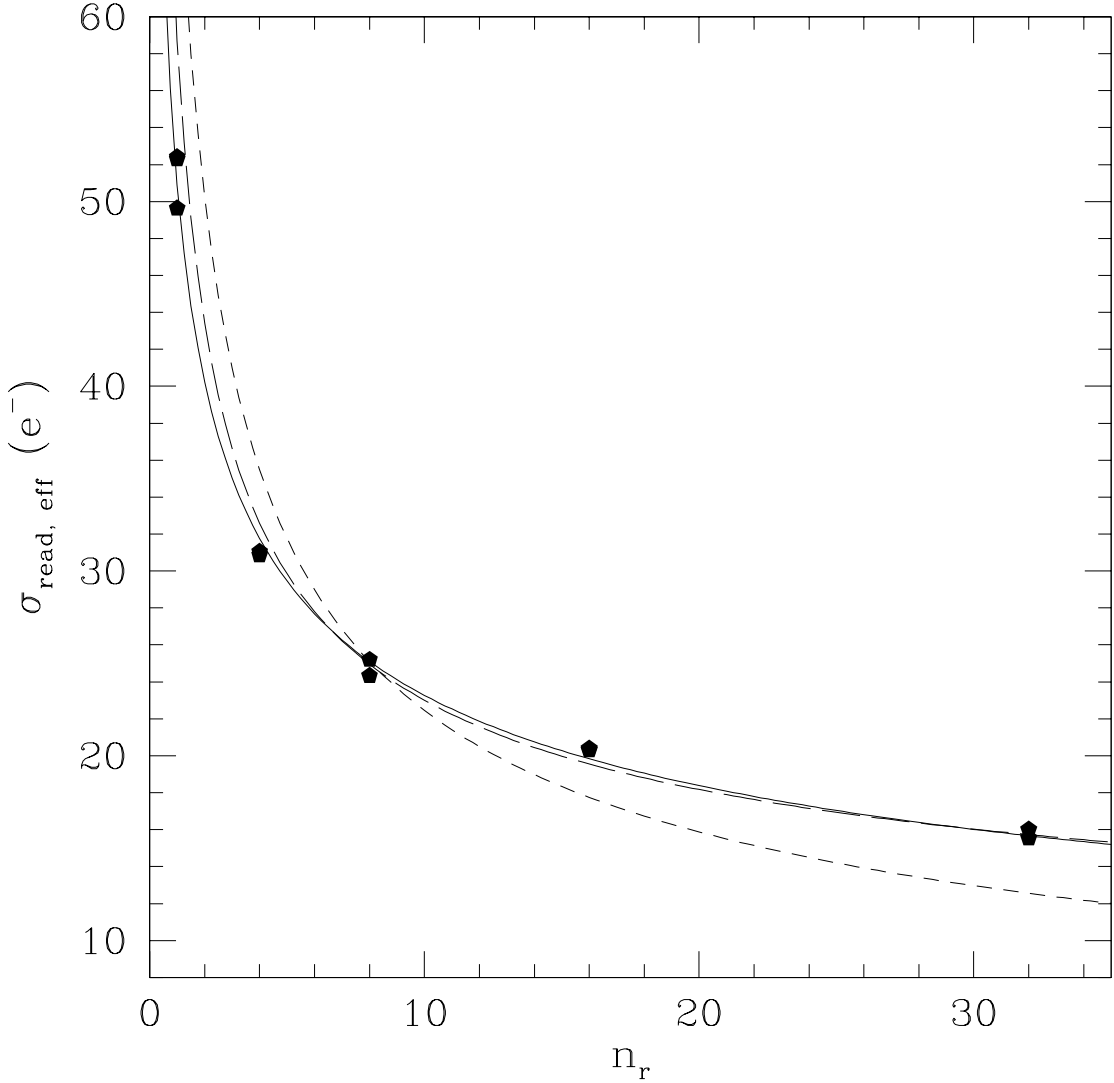,angle=0,height=6in}
\caption{The effective read noise as a function of the number of reads. The solid line is the best fit power law which decreases as $n_r^{-0.34}$, significantly shallower than expected; the short dashed line is the expected decrease as $n_r^{-0.5}$, normalized to the solid fitted curve at $n_r=8$. The long dashed line is the alternative fitting form in which the effective read noise decreases as $n_r^{-0.5}$
plus an additive constant.}
\end{figure}

\end{document}